\documentstyle[11pt]{article}
 
\begin{document}

\title{{\bf V485 Centauri: the Shortest Period SU~UMa 
Star}\footnote{Based on observations obtained with the 1.3~m Warsaw telescope at 
the Las Campanas Observatory of the Carnegie Institution of Washington}} 
 
\author{Arkadiusz~~O~l~e~c~h}

\date{Warsaw University Observatory, 
Al.~Ujazdowskie~4,~00-478~Warszawa, Poland\\
e-mail: olech@sirius.astrouw.edu.pl} 

\maketitle

\abstract{We report CCD observations of the dwarf nova V485~Cen during its 
1997 superoutburst. The long lasting outburst with characteristic slope, 
brightness dip and presence of clear superhumps give evidence that V485~Cen 
belongs to the group of SU UMa-type variables. The superhump 
period is equal to ${0.^d042156\pm0.^d000004}$  
(${60.^m7}$). Thus, V485~Cen has the shortest known 
period among SU UMa-type stars, with a period as much as 25\% shorter than 
previously known objects. The analysis of times of superhump maxima gives 
clear evidence for the increase of superhump period with ${\dot{P}_{\rm 
sh}=28.3\times10^{-5}}$. This fact confirms the hypothesis that the shortest 
periods SU~UMa variables contrary to other stars of this type have positive 
values of ${\dot{P}_{\rm sh}}$.}


\section{Introduction}
Cataclysmic variable stars are binary systems containing white dwarf primary 
and late-type main sequence, low mass secondary. The secondary fills its 
Roche lobe and loses material through the inner Lagrangian point toward the 
white dwarf primary. In case of non-magnetic systems 
falling  material forms a disc around the white dwarf. 

Dwarf novae are a subclass of cataclysmic variable systems (for recent reviews 
see Warner 1995 and Osaki 1996). Usually they are divided into three 
additional classes. The first one is called U~Geminorum (or SS~Cygni) 
type stars. Objects belonging to this type are 
characterized by orbital periods in the range 3--10 hours and by having 
outbursts with an amplitude from 2 to about 8~mag which last a few days, 
separated by a few weeks period of quiescence. The mechanism of such outbursts 
is a thermal instability in the disc which causes episodes of enhanced mass 
transfer from the disc into the primary. The second subgroup are 
Z~Camelopardalis stars. The periods of these objects are in the same range as 
U~Gem-type variables but additionally during the outbursts one can observe 
so-called "standstills". It is believed that Z~Cam stars lay on the border 
line between stars with thermally stable and unstable discs. 

The stars belonging the the third group are called SU~Ursa Majoris systems. 
They have orbital periods in range 80--120 minutes and from time to time they 
show additional, slightly brighter and longer lasting outbursts called 
superoutbursts or super-maxima. A characteristic feature of the superoutbursts 
is presence of "tooth-shape" periodic light oscillations called superhumps. 
The periods of superhumps are 1\%--9\% longer than the binary orbital periods. 
The amplitude of light modulations is typically in the range 0.1--0.4~mag. 

In the Catalog and Atlas of Cataclysmic Variables (Downes and Shara 1993) 
variable star V485~Centauri is classified as U~Geminorum type nova. The 
magnitude range of variability is given as 12.9--17.9~mag. The first 
photometric and spectroscopic study of this star made by Augusteijn et al. 
(1993) revealed clear photometric oscillations with the period 
0.$^d$041096 (59.0~min) and amplitude about 0.3~mag. The spectra presented 
in the same paper were fairly typical for cataclysmic variables and showed 
clear ${\rm H\alpha}$, HeI and CaII emission lines. Due to the presence of 
hydrogen lines Augusteijn et al. (1993) excluded the hypothesis that V485~Cen 
is a double-degenerate AM~CVn system. Also the value of the period was too 
long for the rotational period of the white dwarf primary which is observed in 
the intermediate polars. The main conclusion of that paper was that V485~Cen 
contains hydrogen-deficient main-sequence star and the observed brightness 
oscillations reflect the orbital period of the system. 

The second paper with more extensive quiescent photometry and spectroscopy of 
V485~Cen was published by Augusteijn et al. (1996). They confirmed the 59~min 
periodicity is the orbital period of the system and gave its more exact value 
equal to 0.$^d$040995001. They also gave an estimate of a few 
parameters of the system. The mass ratio $q$ defined as ${M_{\rm WD}}/{M_{\rm 
sec}}$ was estimated to be about 2.6 (lower limit for a mass of the white dwarf 
was ${M_{\rm WD}\approx0.7M_\odot}$ and a lower limit of the mass of the secondary 
${M_{\rm sec}\approx0.14M_\odot}$), the inclination was equal to $i=20-30^\circ$ 
and mass transfer from the secondary ${\dot{M}}$ was estimated between 
${1\times 10^{-10}}$ and ${1\times 10^{-9}~M_\odot/{\rm year}}$. 

In the middle of May 1997 we were notified by Rod Stubbings ({\it VSNET-alert} 
no.~908) that a new outburst of V485~Cen had just begun. In the present paper 
we report on results of CCD photometry of V485~Cen performed during that
outburst. 

\section{Observations and Data Reduction}
The entire set of observations presented in this paper was carried out at Las 
Campanas Observatory in Chile, which is operated by Carnegie Institution of 
Washington. Data were collected with the 1.3~m Warsaw telescope equipped with 
a 2048$\times$2049 SITe thinned CCD with a scale 0.417$"/$pixel. For 
the purpose of observing of V485~Cen we used only part of CCD chip trimming it 
to 512$\times$512 pixels. The detailed description of the system used is given 
by Udalski (1997). 

Observations of V485~Cen were collected as a subproject of the Optical 
Gravitational Lensing Experiment (OGLE-2). The main goal of the OGLE-2 project 
is a search for dark matter in our Galaxy using microlensing phenomena 
(Paczy\'nski 1986, Udalski et al. 1992). When atmospheric conditions are poor 
(seeing $>1.6"$, cirrus clouds) and  photometry of dense stellar regions is not 
reliable some sub-projects like described in this paper are conducted.

We have monitored V485~Cen in $B$, $V$ and $I$ filters on 14 nights from May 
16 through June 2, 1997. The exposure times varied between 60 and 300 seconds, 
depending on  atmospheric conditions and the brightness of the star. Dead 
time between the consecutive frames was about 20 seconds. Journal of 
observations with duration of each run, filters used and exposure times is 
given in Table~1. 

\vspace{0.5cm}
\begin{center}
Table 1 \\
\vspace{0.25cm}
{Journal of the CCD observations of V485~Cen}
\\
\vspace{0.4cm}
\begin{tabular}{|l|c|c|c|c|}
\hline
Date & Time of start & Length of & Filter & Exp. Time \\
1997 & HJD~2450000.~+ & run (h)  &         & (sec) \\
\hline
May 16/17 & 585.5286 & 3.2 & $B$,$V$,$I$ & 60 \\
May 18/19 & 587.7642 & 1.1 & $I$ & 60 \\
May 19/20 & 588.4735 & 4.0 & $V$,$I$ & 60 \\
May 20/21 & 589.4684 & 5.7 & $V$,$I$ & 60 \\
May 21/22 & 590.4893 & 4.3 & $V$,$I$ & 60 \\
May 22/23 & 591.4569 & 5.8 & $I$ & 60,90 \\
May 23/24 & 592.4646 & 5.0 & $V$,$I$ & 60 \\
May 24/25 & 593.4539 & 1.9 & $V$,$I$ & 90,120 \\
May 25/26 & 594.5252 & -$^*$ & $I$ & 180,300 \\
May 27/28 & 596.5169 & 1.2 & $I$ & 180 \\
May 27/28 & 596.6380 & 0.8 & $I$ & 180 \\ 
May 30/31 & 599.5640 & 2.1 & $V$,$I$ & 90 \\
May 31/01 & 600.4396 & 1.3 & $I$ & 90 \\
June 01/02 & 601.4681 & 1.0 & $I$ & 180 \\
June 02/03 & 602.4490 & 1.0 & $I$ & 180 \\
\hline
\noalign{\vskip3pt}
\multicolumn{5}{l}{$^*$ - only 6 measurements made}\\
\end{tabular}
\end{center}
\vspace{10pt}

The data reduction (debiasing and flatfielding) was performed using 
IRAF\footnote{IRAF is distributed by National Optical Observatories, which is 
operated by the Association of Universities for Research in Astronomy, Inc., 
under cooperative agreement with National Science Foundation.} software. The 
profile photometry was done with the DAOphotII package. All data were 
differentially reduced using a comparison star located ${\sim50"}$~W and 
${\sim10"}$~S with respect to the variable star. According to Augusteijn et al. 
(1993) the $V$ brightness of the comparison star is ${15.04\pm0.02}$~mag and 
its color ${B-V=0.64\pm0.04}$. Because our $V$-band is very close to the 
standard Johnson's band we simply added Augusteijn et al. (1993) value to our 
differential measurement to have absolute scale. 

Mean errors during the first stage of outburst were between 0.005 and 0.01 mag 
depending mainly on atmospheric conditions. During a brightness dip and the 
second stage of outburst errors were in the range 0.007--0.030 mag except for 
the night May 25/26 when only six exposures through thick cirrus clouds were 
made and errors were about 0.15~mag. Seeing varied from $1.1"$ on the 
best night to $2.1"$ during the worse run. 

\section{Long-term Behavior of V485~Cen}
Based on observations made by members of the Variable Star section of the 
Royal Astronomical Society of New Zealand, Augusteijn et al. (1993) mentioned 
that about twenty outbursts of V485~Cen were observed with a duration between 
1 and 7 days. 

Long-term behavior of V485~Cen during May 1997 outburst is presented in 
Fig.~1. Rough estimate of the zero point on the magnitude axis is 15.0~mag. 
The first  observations were made on May 17.024~UT. Certainly the outburst 
began a little earlier because the first positive detection of this star in 
outburst was made by Rod Stubbings at 9:16~UT on May 15 ({\it VSNET-alert} 
no.~908). 

During the period May 16/17 -- May 23/24 we observed linear decline of the 
brightness of the star with the rate 0.11~mag/day. Very rapid drop of 
brightness occurred on May 24/25 when the star faded by more than 0.5~mag in 
comparison to previous night. During two further nights brightness reached 
level by more than 3~mag below the maximum. Unfortunately due to bad weather 
conditions we do not have any observations from nights May 28/29 and 29/30. 
After this break we observed the star again on May 30/31 and it was brighter 
over 1.5~mag in comparison to measurements made on May 27/28. It is clear that 
V485~Cen showed about 2~mag brightness dip -- a characteristic feature seen in 
light curves of some SU~UMa stars. 

Since May 30/31 to June 1/2 the decline of brightness was much steeper than 
during the first stage of superoutburst and its rate was equal to 
${\sim1.1}$~mag/day. 

It is clearly visible from Fig.~1 that May 1997 outburst lasted at least 16 
days. In comparison with durations of other outbursts reported by observers 
from New Zealand this time is relatively long. 

All above facts, i.e. long lasting outburst with plateau and the $\sim$2~mag 
brightness dip are the evidences for calling this eruption superoutburst. As we 
will see in the next section another property characteristic for superoutburst 
-- superhumps, was also detected. 

\section{Superhumps}

Fig.~2 presents nightly $I$-band light curves of V485~Cen for eight first
nights. The first run from May 16/17 is separated from other nights by
one night and next seven runs from May 18/19 to 24/25 are consecutive.
The superhumps with their characteristic shape of steeper increase to the
maximum and slower decrease are clearly visible on each night. Their
amplitude is about 0.25~mag on May 16/17 and decreases slowly to 0.1~mag
on May 23/24. During the last night presented in Fig.~2 the amplitude
increases to about 0.15~mag.

To obtain the value of superhump period we have used the Lomb--Scargle 
(Lomb 1976, Scargle 1982) method of Fourier analysis for unevenly spaced data. 
Before the calculation of power spectra we have removed the nightly mean and a 
longer-scale change trend from each individual run. The resulting periodogram 
is shown in Fig.~3. The highest peak is detected at frequency 23.7254 
cycles/day which corresponds to a period ${0.^d04215\pm0.^d00009}$. 
Additionally at frequency 47.443 cycles/day the first harmonic of 
main periodicity is clearly visible. 

We also determined 27 times of maxima of superhumps. 
They are listed in Table~2. The best linear fit to these maxima computed by 
least squares method is given below: 
$$\begin{tabular}{r@{\hspace{2pt}}c@{\hspace{2pt}}r@{\hspace{2pt}}c@{\hspace{2pt}}r@{\hspace{2pt}}l}
HJD$_{\rm Max}$ & = & 2450585.5522   &   +   & 0.042156 & $E$ \\
                &   & ${\pm}$~0.0005 & $\pm$ & 0.000004 &     \\
\end{tabular}
\eqno(1)$$

This value is in very good agreement with the period obtained from Fourier 
transform but its accuracy is much better, so we may conclude that the 
superhump period of V485~Cen is equal to 
${0.^d042156\pm0.^d000004}$ (60.$^m$7). Preliminary value of 
superhump period of V485~Cen, equal to 57.7~min, reported by Olech (1997) was 
based on one observing run only, and therefore turned out to be slightly 
incorrect. Thus, V485~Cen has the shortest known period among SU~UMa-type 
stars. The shortest orbital and superhump periods were previously observed in 
WZ~Sge and AL~Com and were 81.$^m$6 and 82.$^m$3 for WZ~Sge and 
81.$^m$6 and 82.$^m$6 for AL~Com (Patterson et al. 1981, Patterson 
et al. 1996, Howell et al. 1996), respectively. 

Knowing the value of the superhump period we used our $V$-band measurements to 
plot phased $V$-band light curves of V485~Cen for five nights of 
superoutburst. Result is shown in Fig.~4. Phase 0.0 corresponds to 
HJD=2450585.5553. Two cycles are shown for clarity. 

\vspace{0.5cm}
\begin{center}
Table 2 \\
\vspace{0.25cm}
{Times of Superhump Maxima of V485~Cen}
\\
\vspace{0.4cm}
\begin{tabular}{|l|c|c||l|c|c|}
\hline
~~HJD & $E$ & ${O-C}$  &~~~HJD & $E$ & ${O-C}$  \\
2450000.~+ &   & cycles  &2450000.~+  &  & cycles   \\  
\hline
&&&&&\\
   585.5553 &    0 &~~0.0738 & 590.5234 &  118 &--0.0767 \\
   585.5973 &    1 &~~0.0695 & 590.5667 &  119 &--0.0478 \\  
   585.6386 &    2 &~~0.0507 & 590.6096 &  120 &--0.0304 \\
   587.7878 &   53 &~~0.0319 & 590.6512 &  121 &--0.0449 \\
   588.5021 &   70 &--0.0241 & 591.5398 &  142 &~~0.0035 \\      
   588.5444 &   71 &--0.0217 & 591.5800 &  143 &--0.0116 \\  
   588.5870 &   72 &--0.0101 & 591.6214 &  144 &--0.0290 \\
   588.6296 &   73 &~~0.0000 & 592.4685 &  164 &~~0.0652 \\    
   589.4707 &   93 &--0.0478 & 592.5102 &  165 &~~0.0535 \\
   589.5138 &   94 &--0.0246 & 592.5519 &  166 &~~0.0420 \\
   589.5542 &   95 &--0.0666 & 592.5939 &  167 &~~0.0391 \\
   589.5970 &   96 &--0.0521 & 592.6356 &  168 &~~0.0290 \\
   589.6393 &   97 &--0.0478 & 593.4812 &  188 &~~0.0869 \\
   589.6810 &   98 &--0.0579 &          &      &~~       \\
  \hline
\end{tabular}
\end{center}
\vspace{10pt}

The superhump period given in ephemeris~(1) is a mean value averaged 
from eight nights of superoutburst. In Table~2 we also list the ${O-C}$ 
values calculated with the ephemeris~(1). These residuals evidently indicate 
an increase of the period. The quadratic ephemeris obtained as the best least 
squares fit to the same 27 maxima listed in Table~2 is the following: 
$$\begin{tabular}{r@{\hspace{2pt}}c@{\hspace{2pt}}r@{\hspace{2pt}}c@{\hspace{2pt}}r@{\hspace{2pt}}l@{\hspace{2pt}}c@{\hspace{2pt}}@{\hspace{2pt}}l@{\hspace{2pt}}}
HJD$_{\rm Max}$ & = & 2450585.5555   &   +   & 0.042047 & $E$ & $+$   & 5.96$\times10^{-6}E^2$\\
                &   & ${\pm}$~0.0006 & $\pm$ & 0.000012 &     & $\pm$ & 0.63                  \\
\end{tabular}
\eqno(2)$$

The increase of the superhump period is shown in Fig.~5, where we have plotted 
the ${O-C}$ residuals taken from Table~2. The solid line in Fig.~5 presents 
the fit corresponding to the quadratic ephemeris~(2). 

As it was already pointed out by Semeniuk et al. (1997) it might suggest that 
the SU~UMa stars with the shortest orbital periods  exhibit 
increasing superhump periods contrary to the other objects from this group 
which derivative of superhump period is negative (see Patterson et al. 1993). 

\section{Interpulses}
Beginning from night of May 23/24 we observed clear secondary superhumps 
(sometimes called interpulses) located on the light curve of the star between 
maxima of ordinary superhumps. In the upper panel of Fig.~6 we plotted again light 
curve form May 23/24 and marked interpulses by arrows. Additionally in 
the lower 
panel we plotted the light curve from May 23/24 phased with the period 
0.$^d$042156. In both panels interpulses are clearly visible. 

One additional interpulse was also detected on May 24/25, but this was the last 
night before the brightness dip occurred and after that moment interpulses were 
not observed. The moments of maxima of detected interpulses are listed in 
Table~3. 

\vspace{0.5cm}
\begin{center}
Table 3 \\
\vspace{0.25cm}
{Times of Interpulse Maxima of V485~Cen}
\\
\vspace{0.4cm}
\begin{tabular}{|l|c||l|c|}
\hline  
~~HJD & $E$ &~~~HJD & $E$  \\
2450000.~+ &  cycles  &2450000.~+   & cycles   \\
\hline
&&&\\
592.4777 & ~0 &  592.6067 & ~3 \\
592.5227 & ~1 &  592.6480 & ~4 \\
592.5633 & ~2 &  593.4905 & 24 \\
\hline
\end{tabular}
\end{center}
\vspace{10pt}

Similar interpulses were also observed in other SU~UMa stars like SW~UMa 
(Semeniuk et al. 1997) or SU~UMa itself (Udalski 1990). It was suggested 
(Schoembs and Vogt 1980, Warner 1995) that late superhumps in VW~Hyi may 
develop out of such interpulses. However, we did not have possibility to study 
this hypothesis because of the brightness dip, which started on the second 
night in which we observed interpulses. 

\section{Post--Dip Behavior}

We have collected 5 observing runs during and after the brightness dip in 
light curve of V485~Cen. The observations made during the first night, that is 
May 27/28, showed that star faded over 3~mag below its maximum brightness. 
After two nights we detected the star at brightness only 1.4~mag below the 
maximum. Apparently the brightness dip ended and the second stage of 
superoutburst began. This stage lasted only 2 or 3 days because on June 1 the 
star was again over 3~mag below the maximum brightness. 

Fig.~7 presents nightly light curves of V485~Cen from May 27 to June 2, i.e.
during and after the brightness dip. The periodic light oscillations with 
amplitude from 0.05~mag to 0.25~mag are clearly visible. It is obvious that at 
this time light variations showed double humped structure with two humps (one 
with higher amplitude than the other) present at each cycle. 

Again to obtain the value of the period of these variations we have used the 
Lomb--Scargle (Lomb 1976, Scargle 1982) method of Fourier analysis. Before the 
calculation of power spectra we have also removed the nightly mean and a 
longer-scale change trend from each individual run. The resulting periodogram 
is shown in Fig.~8. The highest peak in the power spectrum corresponds to 
frequency 48.782~cycles/day, i.e. ${0.^d020499\pm0.^d00003}$. Due 
to the double structure of light modulations the real value of the period 
should be twice of that. We also detect a peak at 24.352~cycle/day which 
corresponds to ${0.^d04106\pm0.^d00009}$ and which is within errors 
twice the value of 0.$^d$020499. 

From the light curves presented in Fig.~7 we determined 8 times of maxima of 
higher peaks. They are marked by arrows and additionally listed in Table~4. 

\vspace{0.5cm}
\begin{center}
Table 4 \\
\vspace{0.25cm}
{Times of Orbital Maxima of V485~Cen}
\\
\vspace{0.4cm}
\begin{tabular}{|l|c||l|c|}
\hline
~~HJD & $E$ &~~~HJD & $E$  \\
2450000.~+ &  cycles  &2450000.~+   & cycles   \\
\hline
&&&\\
596.5194 & ~0 &  599.6336 & ~76 \\
596.5601 & ~1 &  600.4565 & ~96 \\
596.6425 & ~3 &  601.4783 & 121 \\
599.5937 & 75 &  602.4644 & 145 \\
\hline
\end{tabular}
\end{center}
\vspace{10pt}

The best linear fit to these maxima calculated by the least squares method gives 
the following ephemeris: 
$$\begin{tabular}{r@{\hspace{2pt}}c@{\hspace{2pt}}r@{\hspace{2pt}}c@{\hspace{2pt}}r@{\hspace{2pt}}l}
 
HJD$_{\rm Max}$ & = & 2450596.5192   &   +   & 0.040996 & $E$ \\
                &   & ${\pm}$~0.0006 & $\pm$ & 0.000007 &     \\
\end{tabular}
\eqno(3)$$

\noindent which is in very good agreement with the value of period obtained from Fourier 
spectrum. 

We can summarize that during the five nights from May 27 to June 2 the light 
curve was dominated by double humped structure with a period  
$0.^d0409961\pm0.^d0000066$ (${59.^m0}$). This is in 
excellent agreement with the value of orbital period of V485~Cen given by 
Augusteijn et al. (1996) and equal to $0.^d040995001$. 

\section{Discussion}

Recently a few groups published their results concerning 1995 superoutburst of 
AL~Comae Berenices (Pych and Olech 1995, Kato et al. 1996, Howell et al. 1996, 
and Patterson et al. 1996). The conclusions were in all cases very similar -- 
AL~Com is a twin star of previously well studied dwarf nova WZ~Sge. Both 
objects have one of the shortest known orbital periods among SU~UMa stars, 
both show very rare superoutburst with amplitude about 8--9~mag and almost no 
ordinary outburst. Both also show a 2--3~mag brightness dip in the light 
curve, and in both stars dominant feature during quiescence is very 
symmetrical double hump with the period equal to the orbital period of the 
system and full amplitude of 0.12~mag. Up to now there are only two dwarf 
novae with orbital wave of that shape. Additionally AL~Com during the main 
stage of superoutburst, when ordinary superhumps were clearly visible, showed 
increase of the period of superhumps. There is only one known SU~UMa star, 
except for AL~Com, with positive superhump period derivative. That object is 
SW~UMa (Semeniuk et al. 1997) and it is also characterized by very short 
orbital and superhump periods and long time between superoutbursts. 

Our CCD photometry of 1997 superoutburst of V485~Cen indicates that the star 
belongs to the SU~UMa type variables and shows features very similar to those 
observed in the above mentioned stars. We found that present superoutburst of 
V485~Cen lasted at least 16 days and it is the only one superoutburst known for 
this object. Before, only ordinary outbursts with duration between 1 and 7 
days were observed (Bateson 1979, 1982). 

During the first stage of superoutburst we detected clear periodic light 
modulations called superhumps characteristic for SU~UMa stars. Their period 
was equal to ${60.^m705\pm0.^m006}$. Thus, V485~Cen has the 
shortest known period among SU~UMa type stars, as much as 25\% 
shorter than previously known objects (superhump periods above mentioned stars 
are about 82~min). As it was shown by Paczy\'nski (1981) and Paczy\'nski and 
Sienkiewicz (1981) the theoretical value of minimal orbital period of system 
containing hydrogen-rich secondary is about 81 minutes. That was in very good 
agreement with observational results because the shortest orbital periods of 
SU~UMa stars were about 81~min. Short values of superhump period equal to 
60.$^m$705 and orbital period equal to 59.$^m$03 (Augusteijn et al. 
1996, this work) observed in V485~Cen may suggest that this star belongs to a 
group of AM~CVn systems containing degenerate helium secondary. However, the 
detection of hydrogen emission lines in spectrum of V485~Cen (Augusteijn et al. 
1993, 1996) excludes this possibility. The only remaining possibility is that 
the secondary star in V485~Cen is not degenerate, but a hydrogen deficient 
main-sequence star. Theoretical calculations made by Sienkiewicz (1984), 
Nelson et al. (1986) and Tutukov and Yungelson (1996) imply that depending on 
opacities and hydrogen fraction in the secondary used in calculations the 
limiting value of minimal orbital period is between 60 and 80 min. According 
to Paczy\'nski and Sienkiewicz (1981) absolute minimal orbital period for a 
non-degenerate secondary with critical mass equal to $0.084M_\odot$ is about 49~min. 

The period excess defined as $(P_{\rm sh}-P_{\rm orb})/P_{\rm orb}$ is equal to 
0.028 for V485~Cen. It is known that SU~UMa stars show linear relation between 
the period excess and orbital period with the smallest excesses at short 
orbital periods (Stolz and Schoembs 1981). The periods excesses for the 
shortest period SU~UMa stars like WZ~Sge, AL~Com, HV~Vir and SW~UMa are 0.008, 
0.011, 0.011 and 0.024, respectively. It is clearly visible that V485~Cen does 
not follow the Stolz and Schoembs' relation. According to Fig.~1 of Molnar and 
Kobulnicky (1992) there is another exception from this rule -- T~Leo, which 
period excess is considerably too large for its orbital period. On the other 
hand, calculations made by Whitehurst (1988) show that there is also a linear 
relation between period excess and mass ratio of the system defined as 
${q={M_{\rm sec}/{M_{\rm WD}}}}$. Observational results seem to confirm these 
calculations without any exceptions as can be seen in Fig.~2 of Molnar and 
Kobulnicky (1992). Assuming that V485~Cen is similar to T~Leo in the 
sense that it follows only 
the relation between period excess and $q$ we can roughly estimate its mass 
ratio. From linear relation obtained by Molnar and Kobulnicky (1992), for 
period excess equal to 0.028, $q$ should be around 0.17. This value is in 
disagreement with estimate made by Augusteijn et al. (1996). From spectroscopy 
they obtained $M_{\rm WD}/{M_{\rm sec}}$ about 2.6, that is ${q\approx0.38}$. 
According to Molnar and Kobulnicky (1992 and references therein) the absolute 
upper limit for $q$ for SU~UMa stars is equal to 0.33 and is very likely less 
than 0.22. For higher values of the mass ratio $q$ it is hard to obtain tidal 
instability of accretion disc caused by effect of 1:3 resonance which is 
believed to be a cause of presence of superhumps. Detection of superhumps in 
V485~Cen implies that its mass ratio should be smaller than 0.33. Our rough 
estimate equal to 0.17 is very close to the optimal value ${q=0.16}$ for which 
there is the highest possibility of developement of superhumps (Molnar and 
Kobulnicky 1992). 

We have also demonstrated that the value of the superhump period increases 
during the superoutburst. The period derivative ${\dot{P_{\rm sh}}}$, obtained 
from a parabolic fit to the ${O-C}$ diagram, is equal to ${28.3\times10^{-
5}}$. Such a rate of the superhump period change is more than two times 
greater than the typical value for SU~UMa stars. Moreover the SU~UMa stars 
generally show decreasing superhump periods during the plateau phase of 
superoutbursts. All SU~UMa stars with measured superhump period changes listed 
by Patterson et al. (1993) have negative ${\dot{P}_{\rm sh}}$. The newly 
discovered exceptions from that rule are AL~Com (Howell et al. 1996, Patterson 
et al. 1996) and SW~UMa (Semeniuk et al. 1997) both having the  shortest orbital 
periods among SU~UMa stars and longest intervals between superoutbursts. The 
discovery of positive value of ${\dot{P}_{\rm sh}}$ in V485~Cen confirms 
hypothesis of Semeniuk et al. (1997) that the SU~UMa stars with the shortest 
orbital periods and the longest superoutburst recurrence times exhibit 
increasing superhump periods contrary to the other SU~UMa stars whose 
${\dot{P_{\rm sh}}}$ have negative values. 

The first stage of superoutburst ended with over 2~mag brightness dip which 
lasted 2--3 days. During the  dip and later when star brightened for 2--3 
days we observed double humped light variations with the period 
${59.^m03\pm0.^m01}$, which is exactly  equal to the orbital period 
of the system (Augusteijn et al. 1993, 1996). These modulations were also seen 
after the end of the superoutburst. There are only two other cataclysmic 
variable stars known with orbital waves similar to the observed. These are 
WZ~Sge and AL~Com -- again both having the shortest orbital periods among 
SU~UMa stars and the longest intervals between superoutbursts. 

Kato et al. (1996), Howell et al. (1996), and Patterson et al. (1996) interpreted 
a brightness dip in the light curve of AL~Com as an effect of propagating 
cooling wave in the disc. This cooling wave due to the large amount of 
material which still exists in the disc is not stable and is reflected by 
a heating wave which starts the next normal outburst which subsequently 
triggers the second fainter superoutburst. This hypothesis was partially 
confirmed by detection of superhumps in the post-dip light curve of AL~Com. In 
case of V485~Cen we did not observe superhumps in the post-dip light curve, 
but we detected clear orbital humps. This may suggest that brightening of the 
star after the dip was a normal outburst. Situation is, however, unclear 
because observations of Augusteijn et al. (1993) did not reveal any modulations 
during regular normal outburst and their observations during quiescence also 
did not show any double humped structure. 

Augusteijn et al. (1996) gave an estimate of the mass transfer from the 
secondary in V485~Cen between ${1\times10^{-10}}M_\odot$/y and ${1\times10^{-9}}M_\odot$/y. From theoretical calculations (see 
Osaki 1996 and references therein) we know that such a large values of 
${\dot{M}}$ are characteristic for systems called "permanent superhumpers" -- 
dwarf novae below the "period gap" which are in permanent superoutburst. The 
values of ${\dot{M}}$ obtained for systems such as WZ~Sge, AL~Com or SW~UMa 
are around ${{\dot{M}}\approx2\times10^{-11}}M_\odot$/y, in other words a few times 
smaller than estimate of Augusteijn et al. (1996). According to the fact that 
V485~Cen often shows normal outbursts its mass transfer should be slightly 
higher than in systems like WZ~Sge and AL~Com. This is in good agreement with 
theoretical calculations made by Nelson et al. (1986) who, for system with 
secondary containing 50\% of hydrogen, obtained minimal period equal to 
${P_{\rm orb}=0.^d99}$, mass of the secondary ${M_{\rm sec}=0.057M_\odot}$ 
and mass transfer rate ${{\dot{M}}=0.58\times10^{-10}}M_\odot$/y. 
\vspace{10pt}

\noindent {\bf Acknowledgments}~~ We would like to thank Prof. Andrzej Udalski for his helpful hints, 
reading and commenting on the manuscript. We are also grateful to Prof. 
Janusz Ka{\l}u\.zny for his help with reduction of the raw data. This work was 
partly supported by the KBN grant BW to the Warsaw University Observatory.

\vspace{15pt}
\begin{center}
REFERENCES
\end{center}
\vspace{10pt}

\noindent{Augusteijn, T., van Kerkwijk, M.H., and van Paradijs,
J.},~{1993},~{\it Astron. Astroph.},~{267},~{L55}

\noindent{Augusteijn, T., van der Hooft, F., de Jong, J.A., and van
Paradijs, J},~{1996},~{\it Astron. Astroph.},~{311},~{889}

\noindent{Bateson, F.M.},~{1979},~{\it Publ. Var. Sect. RASNZ},~{7},~{47}

\noindent{Bateson, F.M.},~{1982},~{\it Publ. Var. Sect. RASNZ},~{10},~{13}

\noindent{Downes, R.A., and Shara, M.M.},~{1993},~{\it PASP},~{105},~{127}

\noindent{Howell, S.B., DeYoung, J.A., Mattei, J.A., Foster, G., Szkody, P.
Cannizzo, J.K., Walker, G., and Fierce, E.},~{1996},~{\it Astron. J.},~{111},~{2367}

\noindent{Kato, T., Nogami, D., Baba, H., Matsumoto, K., Arimoto, J.,
Tanabe, K., and Ishikawa K.},~{1996},~{\it PASJ},~{48},~{L21}

\noindent{Lomb, N.R.},~{1976},~{\it Astroph. and Sp. Science},~{39},~{447}

\noindent{Molnar, L.A., and Kobulnicky, H.A.},~{1992},~{\it Astroph. J.},~{392},~{678}

\noindent{Nelson, L.A., Rappaport, S.A., and Jones, P.C.},~{1986},~{\it Astroph. J.},~{304},~{231}

\noindent{Olech, A.},~{1997},~{\it IAU Circ.},~{~},~{6666}

\noindent{Osaki, Y.},~{1996},~{\it PASP},~{108},~{39}

\noindent{Paczy\'nski, B.},~{1981},~{\it Acta Astron.},~{31},~{1}

\noindent{Paczy\'nski, B., and Sienkiewicz, R.},~{1981},~{\it Astroph. J. Letters},~{248},~{27}

\noindent{Paczy\'nski, B.},~{1986},~{\it Astroph. J.},~{304},~{1}

\noindent{Patterson, J., McGraw, J.T., Coleman, L., and Africano,
J.L.},~{1981},~{\it Astroph. J.},~{248},~{1067}

\noindent{Patterson, J., Bond, H.E., Grauer, A.D., Shafter, A.W., and
Mattei, J.A.},~{1993},~{\it PASP},~{105},~{69}

\noindent{Patterson, J., Augusteijn, T., Harvey, D.A., Skillman, D.R., 
Abbott, T.M.C., and Thorstensen, J.},~{1996},~{\it PASP},~{108},~{748}

\noindent{Pych, W. and Olech, A.},~{1995},~{\it Acta Astron.},~{45},~{385}

\noindent{Scargle, J.D.},~{1982},~{\it Astroph. J.},~{263},~{835}

\noindent{Schoembs, R., and Vogt, N.},~{1980},~{\it Astron. Astroph.},~{91},~{25}

\noindent{Semeniuk, I., Olech, A., Kwast, T., and Nale\.zyty,
M.},~{1997},~{\it Acta Astron.},~{47},~{201}

\noindent{Sienkiewicz, R.},~{1984},~{\it Acta Astron.},~{34},~{325}

\noindent{Stolz, B., and Schoembs, R.},~{1981},~{\it IBVS},~{~},~{No. 2029}

\noindent{Tutukov, A., and Yungelson, L,},~{1996},~{\it MNRAS},~{280},~{1035}

\noindent{Udalski, A.},~{1990},~{\it Astron. J.},~{100},~{226}

\noindent{Udalski, A., Szyma\'nski, M., Ka{\l}u\.zny, J., Kubiak, M., and
Mateo, M.},~{1992},~{\it Acta Astron.},~{42},~{253}

\noindent{Udalski, A.},~{1997},~{\it Acta Astron.},~{submitted}

\noindent{Warner, B.},~{1995},~{"Cataclysmic Variable
Stars" (Cambridge) {\it Cambridge University Series} 28}

\noindent{Whitehurst, R.},~{1988},~{\it MNRAS},~{232},~{35}

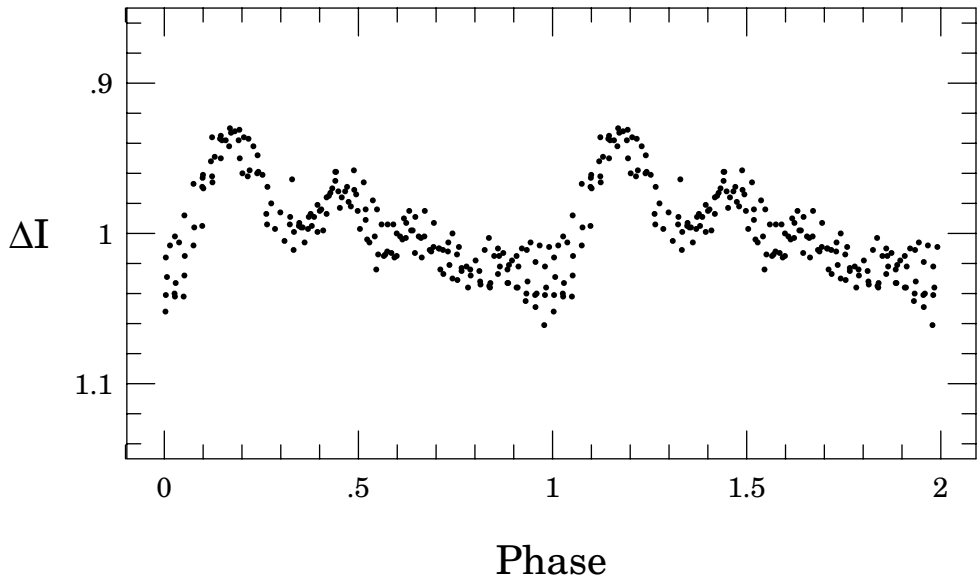
\end{document}